\documentclass[twocolumn,english]{revtex4}
\usepackage{color}
\usepackage{bm}
\usepackage{amsthm}
\usepackage{amsmath}
\usepackage{esint}
\usepackage{graphicx,epsf}
\usepackage{dcolumn}
\usepackage{amsfonts}
\usepackage{mathrsfs}
\usepackage{subfig}
\usepackage{mathrsfs}

\makeatletter
\begin{document}

\title{Soliton generation and drift wave turbulence spreading via geodesic acoustic mode excitation}

\author{Ningfei Chen, Shizhao Wei, Guangyu Wei and Zhiyong Qiu\footnote{E-mail: zqiu@zju.edu.cn}}

\affiliation{Institute for Fusion Theory and Simulation and Department of Physics, Zhejiang University, Hangzhou 310027, China}

\date{\today}

\begin{abstract}
The two-field equations governing fully nonlinear dynamics of the drift
wave (DW) and geodesic acoustic mode (GAM) in the toroidal geometry
are derived in nonlinear gyrokinetic framework. Two stages with distinctive features are identified and analyzed. In the linear growth stage, the set of nonlinear equations can be reduced to
the intensively studied parametric decay instability (PDI), accounting for the spontaneous resonant excitation of GAM by
DW. The main results of previous works on spontaneous GAM excitation, e.g., the much enhanced GAM group velocity and the nonlinear growth rate of GAM, are reproduced from numerical solution of the two-field equations. In the fully nonlinear stage, soliton structures are observed to form due to the balancing of the self-trapping effect by the spontaneously excited GAM and kinetic dispersiveness of DW. The soliton structures enhance  turbulence spreading from DW linearly unstable to stable region, exhibiting convective propagation instead of typical linear dispersive process, and is thus, expected to induce core-edge interaction and nonlocal transport.
\end{abstract}
\maketitle
\section{Introduction}

Anomalous particle and energy transport in magnetically confined fusion devices, generally accepted to be triggered by pressure
gradient-driven drift wave (DW) turbulence \cite{WHortonRMP1999}, is one of the major channels
for thermal plasma transport, and, degradation of plasma confinement.
The size scaling of turbulence transport rate is intensively
studied, however, has not yet been completely understood. The size scaling of turbulent transport obtained from numerical simulations could be significantly
different from theoretical expectation, possibly owing to, e.g., the nonlocality
of transport originate from turbulence spreading \cite{ZLinPRL2002}. More specifically,
the understanding of the mechanism for the turbulence spreading from
linearly unstable to stable region is of fundamental significance,
which may be responsible for the transition from gyro-Bohm scaling
to Bohm scaling and core-edge interaction \cite{HahmPoP2005}.

Radial turbulence spreading due to the toroidal/nonlinear mode coupling was proposed in Ref. \cite{XGarbetNF1994}. Ref. \cite{TSHahmPPCF2004} took a step further,
and adopted a nonlinear reaction-diffusion equation for the turbulence intensity to investigate
a turbulence propagation front radially spreads from linearly unstable
to the stable region with the linear growth and damping rates being the reaction terms, and the nonlinear scattering term being the diffusion term. Despite of the simplified model adopted, the study indicated
that the turbulence spreading contribute to the nonlocality of transport,
i.e., the turbulence intensity can depend on the equilibrium plasma parameters of a distant region, resulting in change of the size scaling of the turbulence intensity and the associated transport. Therefore, the microscopic diffusive turbulence transport
may be intermediated by the meso-scale turbulence spreading process.

On the other hand, spontaneous excitation of zonal field structures (ZFSs)  \cite{FZoncaJPCS2021} including typically electrostatic zonal flows (ZFs), is a significant component of DW nonlinear dynamics. ZFs correspond to mesoscale radial   electric field perturbation with
toroidally symmetric ($n=0$) and  predominantly poloidally  symmetric ($m=0$) mode structures, and consist
of zero-frequency ZF (ZFZF) \cite{ZLinScience1998,LChenPoP2000,PDiamondPPCF2005} and its finite
frequency counterpart, geodesic acoustic mode (GAM) \cite{NWinsorPoF1968,FZoncaEPL2008}.
Here, $n/m$ are the toroidal and poloidal mode numbers of the torus.
It is observed in experiments \cite{KZhaoPRL2006,GConwayPPCF2008,TLanPoP2008,WZhongNF2015,AMelnikovNF2017}
and numerical simulations \cite{ZLinScience1998,YTodoNF2010,XLiaoPoP2016} that
ZFs can be spontaneously excited by DW turbulence including drift Alfv\'en waves (DAWs) and can, in turn, suppress DW turbulence and the associated transport by scattering DW/DAW into stable short radial wavelength domain.

In the existing theoretical investigation, spontaneous excitation of ZFZF/GAM by DW have been studied
in the framework of modulational instability \cite{LChenPoP2000} or parametric decay instability (PDI) \cite{FZoncaEPL2008},
by separating DW into pump wave with fixed amplitude and its sidebands with much smaller amplitudes, and focus on the proof of principle demonstration of ZFZF/GAM excitation by investigating the ``linear''
growth stage of the nonlinear process \cite{LChenPoP2000,FZoncaEPL2008,ZQiuPoP2014}, while the feedback of the small amplitude sidebands to the pump DW is systematically neglected.
This approach, based on separating the DW into pump and small amplitude sidebands, is no longer valid as the amplitudes of
the DW sidebands become comparable to that of the pump wave, i.e., in the ``saturated'' phase as the ``pump" DW is significantly modulated by ZFs. A step forward was made in Ref. \cite{FZoncaPoP2004} where the feedback of DW sidebands and ZFZF
to the pump DW is considered to investigate the pump DW saturation and turbulence spreading. The analysis in Ref. \cite{FZoncaPoP2004} focused on ZFZF induced single-$n$ DW spreading, also embeds the concept of tertiary instability \cite{LChenNF2007a}, which was later extensively studied in, e.g., Ref. \cite{HZhuPRL2020}. Key physics such as higher harmonic generation, however, might still be missing \cite{WWangPoP2006}, which may affect the long time scale dynamics and quantitative saturation level \cite{FZoncaPoP2004}. A two-field model without separating DW into pump and sidebands was thus constructed to investigate the long time scale nonlinear evolution of DW-ZFZF system in the slab geometry \cite{ZGuoPRL2009}, and it was found that coherent soliton
structure was formed, leading to enhanced turbulence spreading. It is worth noting that the soliton structures could also be important for the turbulence spreading due to GAM excitation \cite{FZoncaEPL2008,ZQiuPoP2014}, because the GAM and DW sideband generated during the PDI (the early stage of the nonlinear process) were also shown to couple together and propagate radially due to finite DW/GAM linear group velocities \cite{MRosenbluthPRL1972,ZQiuPoP2014}, but the coupled DW-GAM wave-packet would quickly decouple after propagating out of the pump wave localization region \cite{ZQiuPoP2014}. On the contrary, the soliton structures are characterized by the DW self-trapping, to preserve the shape and amplitude of DW-GAM wave-packet during propagation, and is thus expected to enhance turbulence spreading.

Consequently, in this work, the coupled two-field DW-GAM equations are
derived in the nonlinear gyrokinetic framework with the DW being treated
as a whole. The two-field equations thus include all the necessary ingredients
of the coupled DW-GAM system and will be analysed in this
work to elucidate this potential mechanism for turbulence spreading.
Two phases with distinctive features in the nonlinear process are observed and analyzed, both analytically and numerically. First, in the
early stage of nonlinear evolution, by separating the DW into pump
wave and lower sideband, we show that the two-field model can be
reduced to the intensively studied PDI model \cite{FZoncaEPL2008,ZQiuPoP2014,NChakrabartiPoP2007}. Second, in the later
phase, the soliton structures form as nonlinear coupling induced by the spontaneously excited GAM balances DW kinetic dispersiveness, i.e., DW self-trapping. The nonlinearly generated
soliton structures convectively propagate into a much broader radial extent with $r\sim t$, rather than the typical linear dispersiveness
with $r\sim\sqrt{t}$, which may contribute to the nonlocal transport
and core-edge interplay. Finally, this work offers new insights and
tools into understanding fully nonlinear DW-GAM dynamics.

The rest of the paper is organized as follows. The two-field equations describing
nonlinear interaction of DW and GAM are derived in section \ref{sec:Theoretical-model}.
In section \ref{sec:3}, the reduction of the DW-GAM two-field model in the ``linear'' growth stage to the intensively
studied PDI is presented, both analytically and numerically. We present the major analysis and results
of the two-field model in the fully nonlinear stage in section \ref{sec:4}. Finally, a summary and discussion
of the implication of the DW-GAM soliton structures to the turbulence
spreading are given in the section \ref{sec:5}.

\section{Theoretical model\label{sec:Theoretical-model}}

The derivation of the two-field model for DW-GAM nonlinear interaction
follows closely Refs. \cite{ZQiuPoP2014,FZoncaPoP2004} and \cite{ZGuoPRL2009}.
For simplicity of discussion while focusing on the main physics, it is assumed that the DW and GAM are electrostatic fluctuations,
satisfying proximity to marginal stability, i.e., $|\gamma_{L}/\omega|\ll1$,
such that the parallel mode structure of DW is not disturbed in the nonlinear envelope modulation process.
Besides, each fluctuation is composed of a single $n\neq0$ DW and
GAM with $n=m=0$ scalar potential perturbation. This corresponds to assuming finite but small $\tau k_r\rho_i$, with $\tau\equiv T_e/T_i$ and $k_r\rho_i$ being the radial wavenumber $k_r$ normalized to thermal ion Larmor radius $\rho_i=v_{ti}/\omega_{c,i}$, where $v_{ts}=\sqrt{2T_s/m_s}$ is the thermal velocity and $\omega_{c,s}=e_sB/(m_sc)$ is the cyclotron frequency for species $s$. Inclusion of secondary $m\neq0$ poloidal harmonics of GAM scalar potential is straightforward \cite{ZQiuPPCF2009,ZQiuPST2018}, and will not affect the main results of the analysis. To analyze
the fully nonlinear dynamics, we need to keep DW as a whole, instead
of separating into fixed amplitude pump and lower sideband,
as typically assumed in deriving the PDI dispersion relation \cite{FZoncaEPL2008}. The fluctuations can be expressed as
\begin{eqnarray*}
\delta\phi_{d} & = & A_{d}e^{-i\omega_{d}t-in\zeta}\sum_{m}e^{im\theta}\Phi_{0}(nq-m)+c.c.,\\
\delta\phi_{G} & = & A_{G}e^{-i\omega_{G}t}+c.c.,
\end{eqnarray*}
where subscripts ``d'' and ``G'' denote quantities associated with
DW and GAM, respectively. Furthermore, $A_{d}$ and $A_{G}$ are
the radial envelopes  of DW and GAM, and $\Phi_{0}(nq-m)$ represents
the fine radial structure of DW due to finite $k_{\parallel}$, with
$k_{\parallel}\equiv(nq-m)/qR$ being the parallel wavenumber. For simplicity of discussion while focusing on the main scope of the paper,
a large aspect ratio axisymmetric tokamak with magnetic field given
by $\mathbf{B}=B_{0}[\mathbf{e}_{\xi}/(1+\epsilon\cos\theta)+\epsilon\mathbf{e}_{\theta}/q]$
is considered, where $\epsilon\equiv r/R\ll1$ is the inverse aspect
ratio, $R$ and $r$ are major and minor radii of the tokamak, respectively. Particle response of species
$s$ consists of adiabatic response $e_s\delta\phi F_{0,s}/T_{s}$ and
nonadiabatic response $\delta H_{k,s}$, with $F_{0,s}$ being the
equilibrium distribution function and the subscript ``$k$'' representing ``d'' or ``G''.
Low frequency fluctuations, such as DW and GAM of interest here, can be investigated
using the nonlinear gyrokinetic equation \cite{EFriemanPoF1982}
\begin{eqnarray}
 & \left(\omega-k_{\parallel}v_{\parallel}+\omega_{D,s}\right)\delta H_{k,s}=\dfrac{e_s}{T_{s}}J_{0}\left(k_{\perp}\rho_{L,s}\right)\left(\omega+\omega_{*,s}\right)F_{0,s}\nonumber \\
 & -i\Lambda_{\mathbf{k'},\mathbf{k''}}^{\mathbf{k}}J_{k'}\delta H_{k''}\delta\phi_{k'}.\label{eq:NLGKE}
\end{eqnarray}
Here, $J_{0}\left(k_{\perp}\rho_{L,s}\right)$
is the Bessel function of zero-index representing finite Larmor radius effect (FLR), $\omega_{*,s}= k_\theta cT_s(1+\eta_s(v^2/v_{ts}^2-3/2))/(e_sBL_{n})$ is the diamagnetic
frequency accounting for plasma nonuniformity in radial direction
and $\omega_{D,s}= \mathbf{k}\cdot(\mathbf{B}\times\nabla \mathbf{B}/B^2)(v_\perp^2/2+v_\parallel^2)/\omega_{c,s}$ is the magnetic drift frequency, where $\rho_{L,s}=v_{\perp}/\omega_{c,s}$ is the Larmor radius, $\eta_s\equiv L_{n}/L_{T,s}$ is the ratio of the characteristic length of density variation $L^{-1}_{n}\equiv -\partial\ln(n)/\partial r$ and temperature variation $L^{-1}_{T,s}\equiv-\partial\ln(T_s)/\partial r$. The last
term is the formal nonlinear term, with $\Lambda_{\mathbf{k'},\mathbf{k''}}^{\mathbf{k}}\equiv(c/B_0)\sum_{\mathbf{k}=\mathbf{k'}+\mathbf{k''}}\mathbf{b}\cdot{\left(\mathbf{k''}\times\mathbf{k'}\right)}$ indicating selection
rule for mode-mode coupling, and other notations are standard.

The nonlinear DW-GAM governing equation can be obtained from the charge
quasi-neutrality condition
\begin{eqnarray}
\dfrac{e^{2}N_{0}}{T_{i}}\left(1+\dfrac{T_{i}}{T_{e}}\right)\delta\phi_{k} & = & \left\langle e\delta H_{i}J_{0}\right\rangle _{k}+\left\langle e\delta H_{e}\right\rangle _{k},\label{eq:QN}
\end{eqnarray}
with the angular bracket indicating velocity space integration. Separating the nonadiabatic response into linear and nonlinear
components by taking $\delta H\equiv\delta H^{L}+\delta H^{NL}$, and noting the
$\omega\gg\omega_{tr,i},\omega_{D,i}$ ordering is typically satisfied for both GAM and
DW, with $\omega_{tr,i}=v_{ti}/(qR)$ being the ion transit frequency,the
governing equation for nonlinear DW-GAM system \cite{FZoncaPoP2004}
can be readily cast into

\begin{eqnarray}
 & \dfrac{e^{2}N_{0}}{T_{i}}\left(1+\dfrac{T_{i}}{T_{e}}\right)\delta\phi_{k}-\left\langle e\delta H_{i}^{L}J_{0}\right\rangle _{k}+\left\langle e\delta H_{e}^{L}\right\rangle _{k}\nonumber \\
 & = -i\dfrac{e}{\omega_{k}}\left\langle \Lambda_{\mathbf{k'},\mathbf{k''}}^{\mathbf{k}}\delta H_{e,k''}\delta\phi_{k'}\right\rangle _{k}-\left\langle e\delta H_{e}^{NL}\right\rangle _{k}\nonumber \\
 & -i\dfrac{e}{\omega_{k}}\left\langle \Lambda_{\mathbf{k'},\mathbf{k''}}^{\mathbf{k}}\left(J_{k}J_{k'}-J_{k''}\right)\delta H_{i,k''}\delta\phi_{k'}\right\rangle .\label{eq:governing}
\end{eqnarray}
The terms on the left-hand side are the formally linear terms giving the linear dispersion
relation, while the nonlinear terms are placed on the right-hand side.

For the nonlinear DW equation, the first term on the RHS of Eq. (\ref{eq:governing})
dominates, because the vanishing $\delta H_{e,d}$ and finite $\delta H_{e,G}$
avoid commutative cancellation. The second term on the RHS from nonlinear electron response, is vanishingly small due to the $\omega\ll k_{\parallel}v_{\parallel,e}$ ordering for electron response to DW.  In addition, the third term on the
RHS with finite contribution from FLR effects, is typically $O(k_{r}^{2}\rho_{i}^{2})$ smaller compared to the first
term. Therefore, the nonlinear DW equation can be derived as
\begin{eqnarray}
\omega D_{d}A_{d} & = & \dfrac{ck_{\theta}}{\tau B_{0}}A_{d}E_{G},\label{eq:dispersionDW}
\end{eqnarray}
where $D_{d}$ is the linear DW dispersion relation defined by
\begin{eqnarray}
D_d\equiv1+\dfrac{T_i}{T_e}-\left\langle\dfrac{T_iJ_0\delta H_i^L}{N_0e}\right\rangle,\nonumber
\end{eqnarray}
and $E_{G}\equiv\partial_{r}A_{G}$
is related to the electric field of GAM.

Similarly, for the nonlinear GAM equation,
the first term on the RHS of Eq. (\ref{eq:governing}) vanishes since
$\delta H_{e,d}=0$. The second term due to nonlinear electron response to GAM, has no contribution  due to $\delta H_{e,d}=0$,  corresponding to no source term in the nonlinear gyrokinetic equation for nonlinear electron response to GAM. The third term on the RHS of Eq. (\ref{eq:governing}) can
be rewritten as
\begin{eqnarray*}
&&\Lambda_{\mathbf{k'},\mathbf{k''}}^{\mathbf{k}}(J_{k}J_{k'}-J_{k''})\delta H_{i,k''}\delta\phi_{k'}\\
& = &\mathbf{b}\cdot\left(\mathbf{k''}\times\mathbf{k'}\right)(J_{k'}-J_{k''})(\delta H_{k''}\delta\phi_{k'}+\delta H_{k'}\delta\phi_{k''})\\
&+&\mathbf{b}\cdot\left(\mathbf{k''}\times\mathbf{k'}\right)(J_{k}-1)(J_{k'}\delta H_{k''}\delta\phi_{k'}-J_{k''}\delta H_{k'}\delta\phi_{k''}).
\end{eqnarray*}
The first term in the above formula dominates over the second by $O(1/(k^2_{\perp}\rho^2_i))$ \cite{FZoncaPoP2004}. Consequently,
the nonlinear GAM equation can be readily derived as
\begin{eqnarray}
\omega\mathscr{E}_G E_{G} & = & -\alpha_{i}\dfrac{ck_{\theta}}{B_{0}}\left(A_{d}\partial_{r}^{2}A_{d}^{*}-c.c.\right).\label{eq:dispersionGAM}
\end{eqnarray}
Here, $\mathscr{E}_G$ is the linear GAM dispersion relation defined by
\begin{eqnarray}
\mathscr{E}_G\equiv1+\dfrac{T_i}{T_e}-\left\langle\dfrac{T_iJ_0\delta H_{i,G}^L}{N_0e\delta\phi_G}\right\rangle+\left\langle\dfrac{T_i\delta H_{e,G}^L}{N_0e\delta\phi_G}\right\rangle,\nonumber
\end{eqnarray}
and $\alpha_{i}$ is an order unity coefficient \cite{LChenPoP2000}. Note that
contribution of energetic particles can be formally included in $\mathscr{E}_G$, to account for the nonlinear interaction of energetic particle induced GAM (EGAM) \cite{RNazikianPRL2008,GFuPRL2008,ZQiuPPCF2010} with DW, which was proposed as an active control method for DW turbulence \cite{ZQiuPST2018} and is under investigation \cite{DZarzosoPRL2013,NChenPoP2021}.
However, in this work, effects of energetic particles will not
be included, to focus on the nonlinear interaction of DW with GAM.
Following this strategy, a simplified expression for the electron DW
dispersion relation with quadratic dispersiveness is assumed as $\omega D_{d}=\omega-\omega_{*}+C_{d}\omega_{*}k_{r}^{2}\rho_{i}^{2}$, while the GAM dispersion relation is
 $\mathscr{E}_G=-1+\omega_{G}^{2}/\omega^{2}+C_{G}\omega_{G}^{2}k_{r}^{2}\rho_{i}^{2}/\omega^2$ \cite{ZQiuPPCF2009}, with $\omega_*\equiv k_\theta cT_i/(eBL_{n})$ being the ion diamagnetic frequency. Substituting the linear dispersion relations into Eqs. (\ref{eq:dispersionDW}) and (\ref{eq:dispersionGAM}),
the coupled nonlinear DW-GAM two-field equations can
be formally written as
\begin{eqnarray}
 & \left(\partial_{t}+\gamma_{d}+i\omega_{*}(r)+iC_{d}\omega_{*}\rho_{i}^{2}\partial_r^2\right)A_{d}=\dfrac{-ick_{\theta}}{\tau B_{0}}A_{d}E_{G},\label{eq:twofieldDW}\\
 & \left(\partial_{t}^{2}+2\gamma_{G}\partial_{t}+\omega_{G}^{2}-C_{G}\omega_{G}^{2}\rho_{i}^{2}\partial_r^2\right)E_{G}\nonumber\\
 & =i\alpha_{i}\dfrac{ck_{\theta}}{B_{0}}\partial_{t}\left(A_{d}\partial_{r}^{2}A_{d}^{*}-c.c.\right).\label{eq:twofieldGAM}
\end{eqnarray}
Here, $C_{d}$ and $C_{G}$ are originated from the kinetic dispersiveness effects of DW and GAM, as derived in Refs. \cite{FRomanelliPoFB1993}
and \cite{ZQiuPPCF2009}, respectively. Furthermore, $\gamma_{d}$ and $\gamma_{G}$
are the linear growth/damping rates of DW and GAM, respectively.  The RHS of the two-field equations are the nonlinear terms, indicating the modulation of DW envelope by GAM, and spontaneous excitation of GAM by Reynolds stress from DW nonlinearity. For clarity of discussion,
the linear growth/damping rates $\gamma_{d}$, $\gamma_{G}$ and system
nonuniformity (dependence of $\omega_*(r)$ on $r$) are neglected in the following analysis, to focus on the nonlinear interaction process.
The analytical solution of the coupled DW-GAM two-field equations can be obtained
in some limited cases, however, the equations are mainly solved numerically in the rest of the present work,
which can also provide necessary clues for theoretical treatment,
as presented in the following. For the convenience of numerical investigation,
space and time can be normalized to $\rho_{i}$ and $\omega_{*}^{-1}$, respectively,
i.e., $r\rightarrow r/\rho_{i}$, $t\rightarrow\omega_{*}t$, $k_{r}\rightarrow k_{r}\rho_{i}$, $\omega\rightarrow\omega/\omega_{*}$, and $A_{d}$ and $E_{G}$ are normalized to $e/T_{e}$. Thus, the normalized DW-GAM two field equations can be cast into
\begin{eqnarray}
 & \left(\partial_{t}+i+iC_{d}\partial_r^2\right)A_{d}=-i\Gamma_{0}A_{d}E_{G},\label{eq:normalizedDW}\\
 & \left(\partial_{t}^{2}+\dfrac{\omega_{G}^{2}}{\omega_{*}^{2}}-C_{G}\dfrac{\omega_{G}^{2}}{\omega_{*}^{2}}\partial_r^2\right)E_{G}\nonumber \\
 & =i\alpha_i\tau\Gamma_{0}\partial_{t}\left(A_{d}\partial_{r}^{2}A_{d}^{*}-c.c.\right).\label{eq:normalizedGAM}
\end{eqnarray}
Here, $\Gamma_{0}\equiv(T_{i}/e)ck_{\theta}/(B_{0}\rho_{i}\omega_{*})=L_{n}/(2\rho_{i})$
is the nonlinear coupling  coefficient. Note that $\Gamma_{0}$
is related to $\rho_{i}/L_{n}\sim\rho_{i}/a\equiv\rho_{*}$, i.e.,
turbulence intensity size scaling.

In particular, the numerical scheme
used is spectral method, which is accomplished by Fourier transformation in radial direction and formal forth order Runge-Kutta method to advance in time.
The nonlinear terms, treated as convolution sum for formal spectral
method, are otherwise multiplied in physical space and transformed
back to Fourier space in this work to save computing power. This method
is the so called \textit{pesudospectral} method. Finally, we adopt the outgoing boundary
condition to avoid un-physical reflections back to the computation
region, by imposing an artificial dissipation layer near the boundary.

\section{Reduction to the three-wave interaction\label{sec:3}}

\begin{figure}
\includegraphics[scale=0.45]{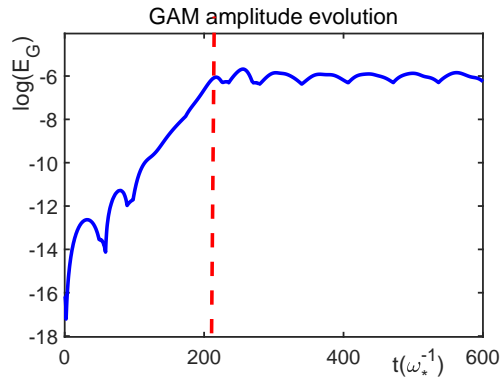}
\caption{\label{fig:1} Evolution of GAM amplitude.}
\end{figure}

There are two phases with distinctive features in the nonlinear evolution of the coupled DW
and GAM system, as shown in Fig. \ref{fig:1}, where the dependence of GAM amplitude on time is given. Specifically, the first phase,
occurring in $t=0-200\omega_{*}^{-1}$ in Fig. \ref{fig:1}, shows an exponential growth of
GAM as a result of the spontaneous excitation by DW \cite{FZoncaEPL2008,ZQiuPoP2014}, while
the second phase, dominating in $t=200-600\omega_{*}^{-1}$, corresponds a saturated
stage characterised by comparable level of DW and GAM, oscillating in a manner of a limited cycle oscillation.
The first phase corresponds to the ``linear''
growth stage of the nonlinear process, typically referred to as PDI phase, and has been studied intensively \cite{FZoncaEPL2008,NChakrabartiPoP2008,ZQiuPoP2014}. Hence, before proceeding
with the more complicated fully nonlinear process, we shall demonstrate the validity and generality of the present two-field model by recovering the results obtained from the well-studied PDI model in the early stage of the nonlinear process.

  In the ``linear'' growth stage where the DW is only slightly modified by nonlinear effects,
DW can be separated into a pump wave ($\omega_{P},k_{P}$) with fixed amplitude and a lower
sideband ($\omega_{S},k_{S}$) with the amplitude much smaller than that of the pump DW, i.e., $\phi_{d}=\phi_{P}+\phi_{S}^{*}$
 with the subscripts ``P'' and ``S'' denoting pump DW and sideband, respectively.
This corresponds to the pump DW (mother wave) resonantly pumps energy to two daughter waves, i.e., DW sideband
and GAM; while on the other hand, the feedback of two daughter waves to the pump DW is negligibly small.

The DW lower sideband equation is readily obtained from Eq. (\ref{eq:normalizedDW})
with selection rule in frequency/wavenumber matching conditions applied
\begin{eqnarray}
\omega_{S}D_{d}A_{S} & = & \Gamma_{0}A_{P}^{*}E_{G}.\label{eq:DWPDI}
\end{eqnarray}
Similarly, the GAM equation is obtained from Eq. (\ref{eq:normalizedGAM}) as
\begin{eqnarray}
\omega_{G}^{2}\mathscr{E}_GE_{G}  & = & -\alpha_{i}\tau\Gamma_{0}A_{P}\partial_{t}\partial_{r}^{2}A_{S}.\label{eq:GAMPDI}
\end{eqnarray}
Eqs. (\ref{eq:DWPDI}) and
(\ref{eq:GAMPDI}) are exactly Eqs. (9) and (10) of Ref. \cite{ZQiuPoP2014}, with the resonant $k_{r}$ and $\omega$ obtained from the frequency and wavenumber matching conditions, i.e., $\omega-\omega_{P}+\omega_{*}-C_{d}\omega_{*}\rho_{i}^{2}k_{r}^{2}=0$,
and $\omega^{2}-\omega_{G}^{2}-C_{G}\omega_{G}^{2}\rho_{i}^{2}k_{r}^{2}=0$.
Thus, it is  verified theoretically that the two-field model
can be reduced to PDI in the proper limit.

As obtained in previous investigation \cite{ZQiuPoP2014}, there are several main conclusions
revealed  from the PDI, which are, first, GAM is excited as the nonlinear drive from DW overcomes the threshold due to GAM and DW Landau damping rates; second, the nonlinear growth rate is proportional
to $k_{r}\Gamma_{0}A_{P}$ as the nonlinear drive is well above threshold, which readily gives that excitation of short wavelength kinetic GAM (KGAM) is preferred; and third, the nonlinearly excited GAM and DW lower sideband are coupled together, and propagate at $V_{C}=(V_{G}+V_{d})/2$, with $V_{d}=2C_{d}k_{r}$
and $V_{G}=C_G\omega_Gk_r/\omega_*$ being
the linear group velocities of DW and GAM, respectively. Note that the
linear group velocity of GAM is typically much smaller than that of
DW by $O(\omega_{G}/\omega_{*})$, and that GAM is typically nonlinearly excited by DW turbulence, this explains the experimental
\cite{DKongNF2013} and simulation \cite{RHagerPoP2012,RHagerPRL2012}
observations of GAM propagation much faster than that predicted by linear group velocity.
Note again that $k_{r}$ and $\omega$ are determined by the matching condition, and thus the frequency ratio $\omega_G/\omega_*$.

\begin{figure}

\subfloat{\includegraphics[scale=0.3]{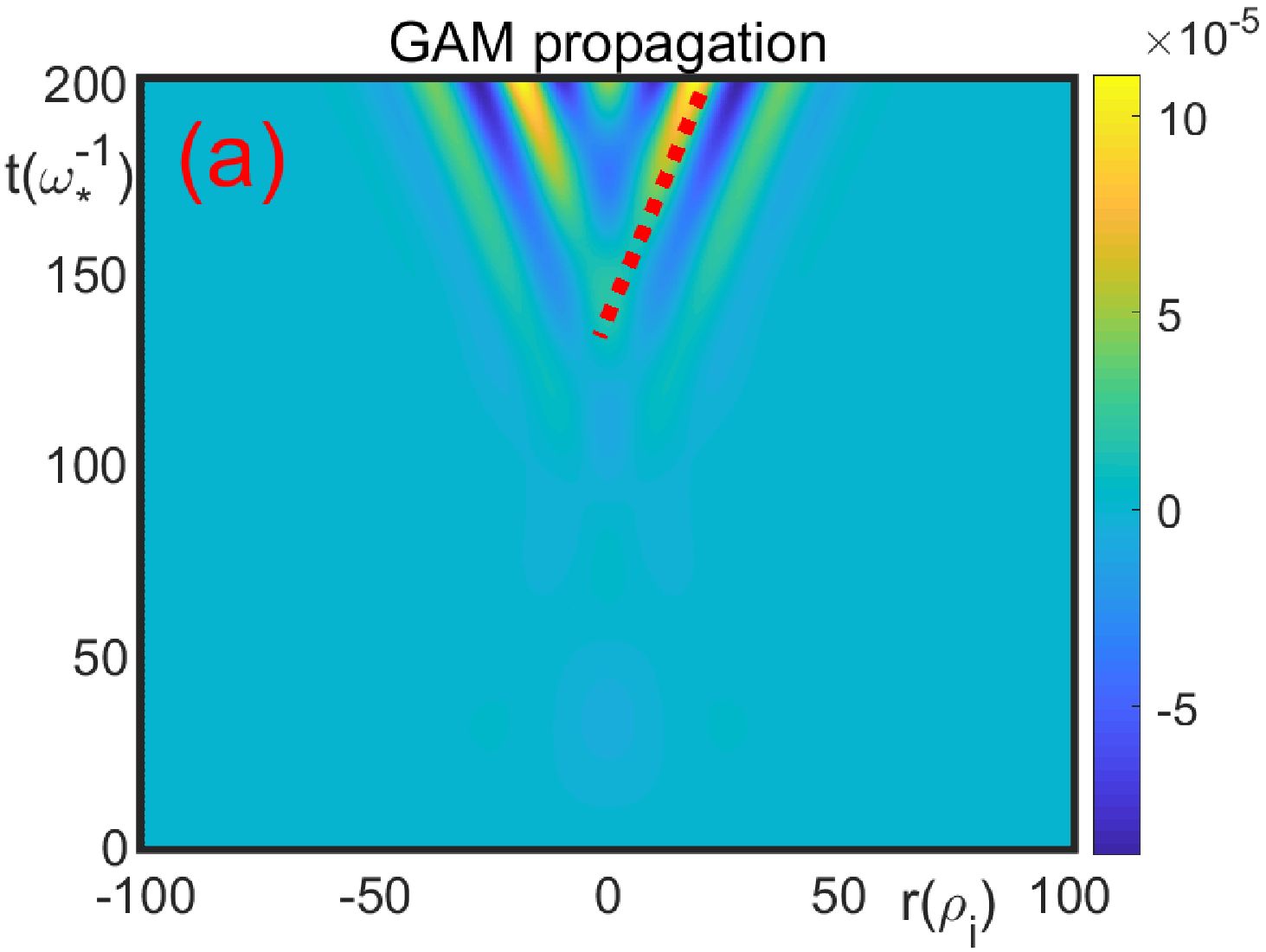}\label{fig:2a}}
\subfloat{\includegraphics[scale=0.3]{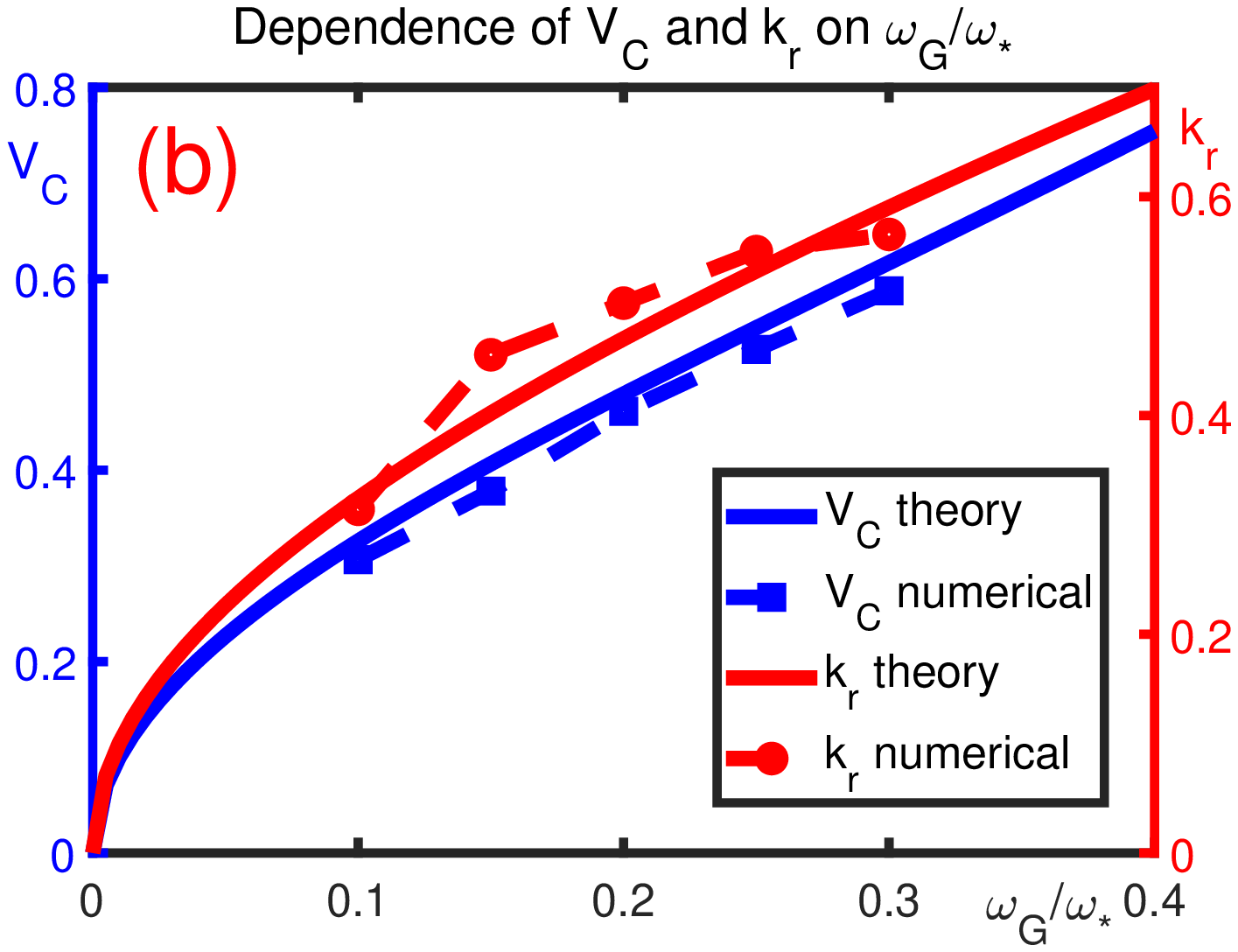}\label{fig:2b}}

\caption{ (a) temporal-spatial evolution of GAM in the linear
growth stage, with the red dashed line indicating the path of the
excited mode; (b) dependence of the group velocity and resonant
$k_{r}$ on frequency ratio $\omega_{G}/\omega_{*}$, obtained both
numerically (dashed) and theoretically (solid).}
\end{figure}

\begin{figure}
\includegraphics[scale=0.45]{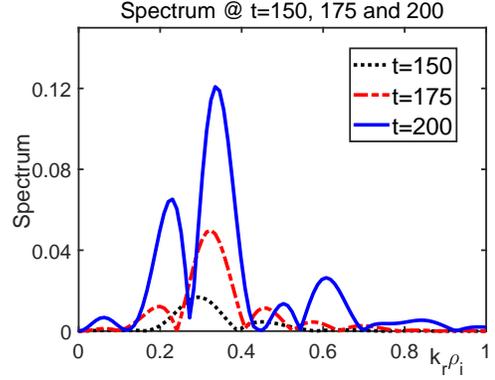}
\caption{The $k_{r}$-spectrum of GAM at $t=150,175,200\omega_{*}^{-1}$, with the frequency ratio $\omega_{G}/\omega_{*}=0.1$.\label{fig:3}}
\end{figure}

These results are reproduced by solving the general fully nonlinear two-field equations numerically.
The spatial-temporal evolution of
GAM is shown in Fig. \ref{fig:2a}, with the red dashed line indicating the path of the spontaneously
excited GAM, from which the GAM nonlinear propagation velocity can be measured, and is compared with  the predicted speed of the coupled DW sideband-GAM
envelope $V_C$ \cite{ZQiuPoP2014}. The dependence of resonant $k_{r}$ and the velocity of the coupled DW sideband-GAM wavepacket on the frequency ratio $\omega_{G}/\omega_{*}$ are given in Fig. \ref{fig:2b}, where the red solid curve is the resonant $k_r$ from matching conditions, and the red dashed curve is the $k_r$ from numerical solution of the two field equations. On the other hand, the theoretically predicted propagation velocity $V_C$ is given by the solid blue curve, while the numerically measured propagation velocity is given by the blue dashed line. The numerically obtained $k_r$ and propagation velocity both agree well with the theoretical solution \cite{ZQiuPoP2014}. In  Fig. \ref{fig:3}, it is shown that, for given $\omega_{G}/\omega_{*}=0.1$, after a transient process,
the resonant radial wavenumber is approximately $k_{r}\rho_{i}=0.32$, which is the same as that obtained from the matching conditions (Fig. \ref{fig:2b}).

\begin{figure}
\subfloat{\includegraphics[scale=0.3]{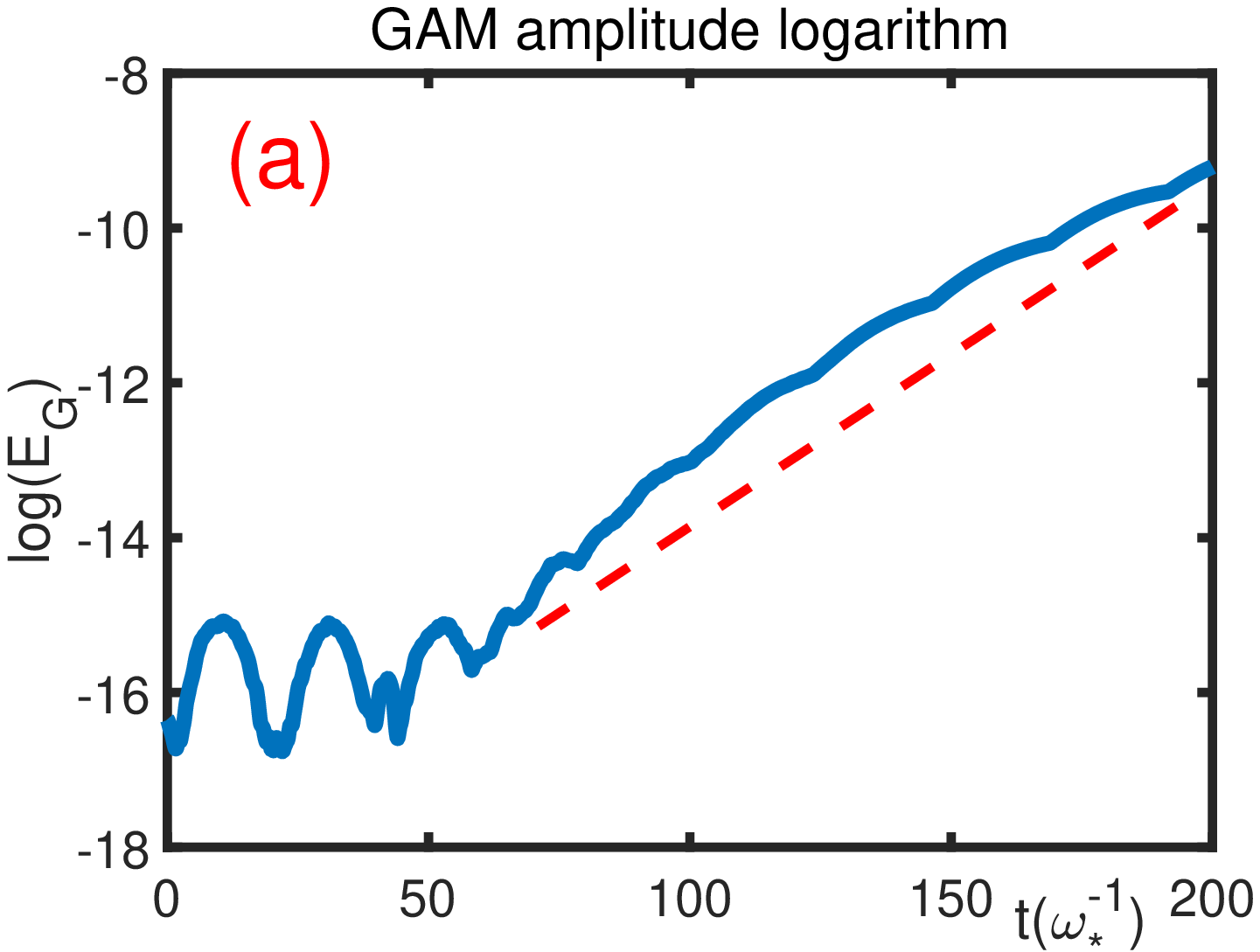}\label{fig:4a}}
\subfloat{\includegraphics[scale=0.3]{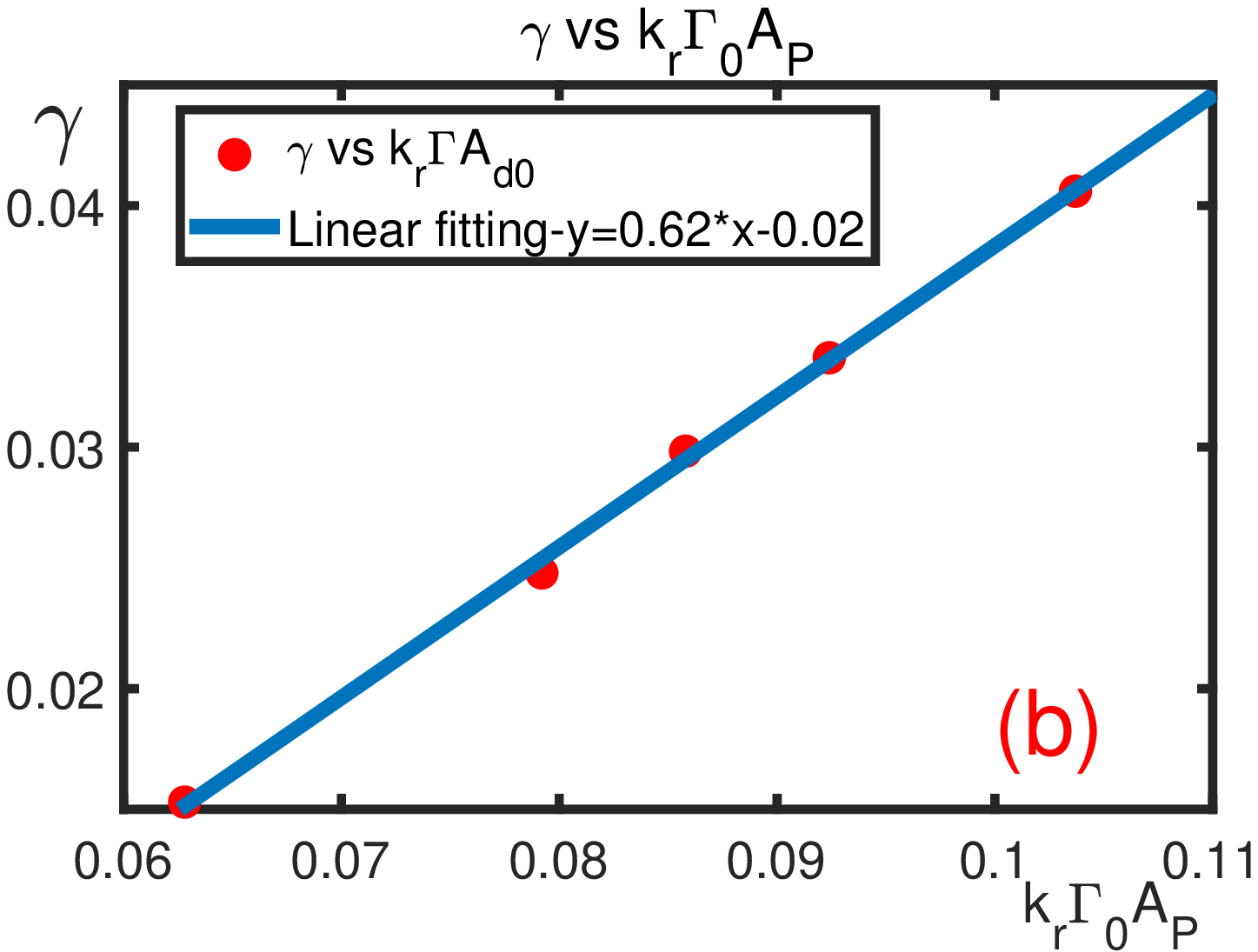}\label{fig:4b}}
\caption{(a) blue solid line is the logarithm of GAM amplitude in $t=0-200\omega_{*}^{-1}$,
while the red dashed line indicates the linear fitting; (b) dependence of PDI growth rate on the $k_{r}\Gamma_0A_{P}$, with the red dots representing the numerical results and the blue solid line being the linear fitting.}
\end{figure}

As for the nonlinear growth rate of GAM, as is shown in Fig. \ref{fig:4a}, the GAM amplitude grows quasi-exponentially with time, with its growth rate being measured from the slope of the red dashed line.  The GAM growth rate dependence on $k_{r}\Gamma_{0}A_{P}$ is plotted in Fig. \ref{fig:4b}, showing a clear linear dependence,
which indicates that the excitation of shorter radial wavelength KGAM is favored. In conclusion, the main results of PDI are convincingly reproduced using the two-field model equations both theoretically and numerically, thus, it would be safe
to conclude that our two-field model is able to recover the PDI in the linear growth stage.

\section{  Soliton formation and  Turbulence spreading\label{sec:4}}

In the second stage of the nonlinear process as shown in Fig. \ref{fig:1} that, the $A_{G}\ll A_{d}$ condition is no longer satisfied (and thus the ``DW sideband"), and both of their amplitudes
saturate, the separation of DW into pump wave and lower sideband is no longer
valid. Therefore, the fully nonlinear DW-GAM two-field model is needed to
account for the higher $k_r$ mode excitation, soliton structure formation by DW self-trapping
and subsequent turbulence spreading \cite{ZGuoPRL2009}.

\begin{figure}

\includegraphics[scale=0.45]{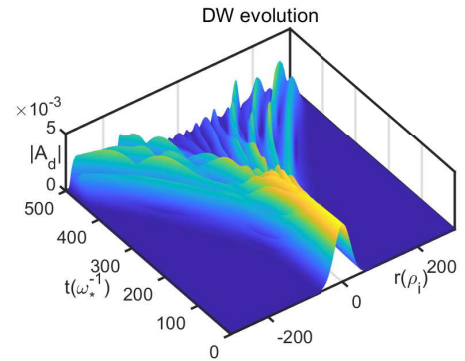}

\caption{Spatial-temporal evolution of DW envelope.\label{fig:5}}

\end{figure}

\begin{figure}

\includegraphics[scale=0.45]{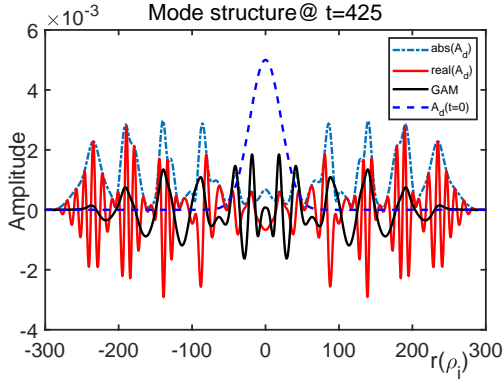}
\caption{The mode structure of DW-GAM envelope soliton at $t=425\omega_{*}^{-1}$, with the deep blue dashed line being the DW initial envelope, blue dashed, red solid and black solid lines being the DW envelope, real part of the DW and GAM, respectively.\label{fig:6}}

\end{figure}

The spatial-temporal evolution of DW envelope obtained from the numerical solution of the coupled two field equations is shown in Fig. \ref{fig:5}, where an initially loaded  DW envelope splits into multiple convectively propagating wave structures after $t=200\omega_{*}^{-1}$ due to spontaneously excited GAM, suggesting the formation of DW solitons \cite{ZGuoPRL2009}.
A snapshot of the DW
and GAM radial mode structures at $t=425\omega_*^{-1}$ is given in Fig. \ref{fig:6}, where one-to-one correspondence of radially fast varying DW structures and slowly varying GAM are clearly exhibited, indicating that the DW radial envelope is
modulated by spontaneously excited GAM radial electric field, i.e., self-modulation/trapping of DW. \footnote{Note that, the observed radial variation of DW turbulence structure in Fig. \ref{fig:6}, due to the modulation by the spontaneously excited zonal structures (not limited to GAM as investigated in the present work), may contribute to yield the first-principle-based interpretation of the  ``staircases" observed from  numerical simulations as well as modelings \cite{GPradalierPRL2015}.}  Consequently, a two scale spatial-temporal analysis can
be employed, by taking $A_{d}=A_{d0}u_{d}\exp(-i\omega t+ik_{r}r)$,
i.e., $\partial_{t}=\partial_{\tau}-i\omega$, $\partial_{r}=\partial_{\xi}+ik_{r}$,
with $k_{r}\gg -i\partial_{\xi}$ and $\omega\gg i\partial_{\tau}$.
Here, $A_{d0}$, $u_{d}$ are the initial amplitude and radial envelope of
the DW, respectively. The normalized DW-GAM two-field Eqs. (\ref{eq:normalizedDW}),
(\ref{eq:normalizedGAM}) can be reduced to
\begin{eqnarray}
& & \left(\partial_{\tau}-i\delta_{d}-V_{d}\partial_{\xi}+iC_{d}\partial_{\xi}^{2}\right)u_{d}=-i\Gamma_{0}u_{d}E_{G},\\
& & (\partial_{\tau}+V_{G}\partial_{\xi})E_{G}=k_{r}\Gamma_{0}\alpha_{i}\tau A_{d0}^{2}\partial_{\xi}u_{d}^{2}.\label{eq:13}
\end{eqnarray}
Here, $\delta_{d}\equiv\omega-1+C_{d}k_{r}^{2}$
is the small deviation from DW linear dispersion relation due to nonlinear effects. Thus, we introduce $\eta=\delta(\xi+V_d\tau)$ to transform the solution $f(\xi,\tau)$ to the soliton frame of reference, in which the soliton solution can be represented as $f(\eta)$, with $\delta$ being a small parameter representing slowly varying radial envelope. Subsequently, the nonlinear equations for the DW and GAM can be combined into a single ordinary differential equation in the form of nonlinear schordinger equation
\begin{eqnarray}
C_{d}\delta^{2}u_{d}''-\delta_{d}u_{d}+\lambda u_{d}^{3} & = & 0.
\end{eqnarray}
Here, $\lambda\equiv k_{r}\alpha_i\tau\Gamma_{0}^{2}A_{d0}^{2}/(V_{d}+V_{G})$ represents
nonlinear trapping effect, and $u_d''=d^2u_d/d\eta^2$. The equation clearly reveals the competition
between the nonlinear trapping and the linear dispersion, while solitary
structures correspond to equating the three coefficients of $u_{d}''$,
$u_{d}$ and $u_{d}^{3}$, that is,
\begin{eqnarray}
 & \delta^{2}C_{d}= & \omega-1+C_{d}k_{r}^{2}=\lambda/2. \label{eq:balancing}
\end{eqnarray}
The nonlinear DW frequency can be expressed as $\omega=1-C_{d}k_{r}^{2}+\lambda/2$, with the last term on the RHS being the nonlinear frequency upshift. The small parameter can also be obtained, $\delta\simeq \Gamma_0 A_{d0}/(2C_d)$. Furthermore, the velocity of the DW soliton can be readily
obtained as $V^{s}_{d}=2C_{d}\sqrt{k_r^2-\lambda/(2C_d)}$, with the superscript ``$s$" denoting soliton, which is valid if $k_r\gtrsim\Gamma_0A_{d0}/{2C_d}\simeq\delta$. This fact is consistent with the assumption of two scale analysis.
The nonlinear velocity deviates from the linear DW group velocity due to the upshift of frequency and
so does the corresponding $k_{r}$ with the same frequency. The DW soliton structure can be solved as $A_{d}=A_{d0}\sinh[\delta(r+V^{s}_{d}t)]\exp(-i\omega t+ik_r r))$, which can be substituted into Eq. (\ref{eq:13}), and yields the GAM electric field as
$E_{G}=E_{G0}\sinh^{2}[\delta(r+V^{s}_{d}t)]\exp(-i\omega t+ik_rr)$,
which is in the hyperbolic sine form, acting as a potential well
to trap the DW envelope, qualitatively consistent with the numerical results.
It is noteworthy that in DW self-trapping region, e.g., $r=150-200\rho_{i}$ in Fig. \ref{fig:6},
GAM electric field can be understood as potential well in the radial direction to trap the DW eigen-function,  in addition to
that of system nonuniformity in Eq. (\ref{eq:twofieldDW}), i.e., the diamagnetic well $\omega_*(r)$.

\begin{figure}
\includegraphics[scale=0.45]{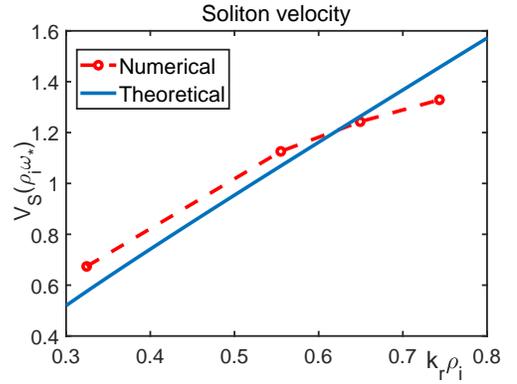}
\caption{ Dependence of  soliton velocity on $k_{r}\rho_{i}$, with the red dashed and blue solid lines representing the numerical and theoretical results, respectively.\label{fig:7}}
\end{figure}

It is crucial to identify the velocity of these solitons due to its relation to nonlocal
transport. The dependence of
the velocity of the DW soliton on $k_{r}$ is shown in Fig. \ref{fig:7}, with the solid curve being the numerical results while the dashed curve corresponding to $V^s_d$. Thus, the DW soliton propagates at
$V^{s}_{d}\propto2k_{r}$ rather than $V_{C}\propto k_{r}$ predicted by the PDI model, due to the self-consistent  trapping by GAM generation.

\begin{figure}
\includegraphics[scale=0.45]{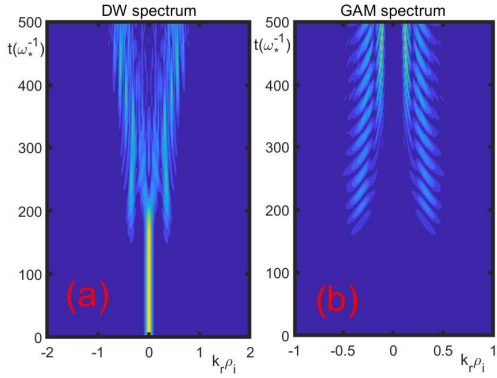}
\caption{The $k_{r}$-spectrum evolution of (a) DW, and (b) GAM for $\omega_G/\omega_*=0.1$, respectively. The matching condition gives $k_r\rho_i\simeq0.32$.\label{fig:8}}
\end{figure}

Besides the evolution of the nonlinear DW-GAM system in physical space, the $k_r$ spectrum evolution is also informative. The left and right panels of Fig. \ref{fig:8} give the $k_r$ spectrum of DW and GAM, respectively. As shown in the left panel of Fig.
\ref{fig:8}, the $k_r$ spectrum of DW is initially a Gaussian centered at $k_r=0$. Later, the parametric excitation of GAM scatters DW into the resonant $k_r$ determined by the matching condition ($k_r\rho_i\simeq 0.32$), on which the subsequent increasing of DW wavenumber is based. The DW is further scattered by GAM into higher $k_r$ regime, as a  result typical for the nonlinear dynamics. The DW dispersiveness is directly related to $k_r$, which competes with the nonlinear trapping effect induced by GAM. However, the $k_{r}$ of excited modes will not increase consistently as expected, possibly due
to the fast propagation of high $k_{r}$ modes out of the interacting zone.

\begin{figure}
\includegraphics[scale=0.45]{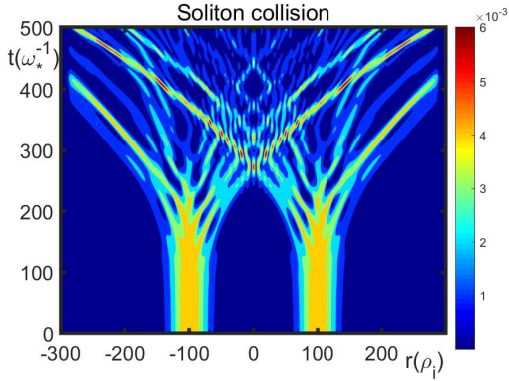}
\caption{ Collision of ``DW'' envelope solitons.\label{fig:9} }
\end{figure}

\begin{figure}
\includegraphics[scale=0.45]{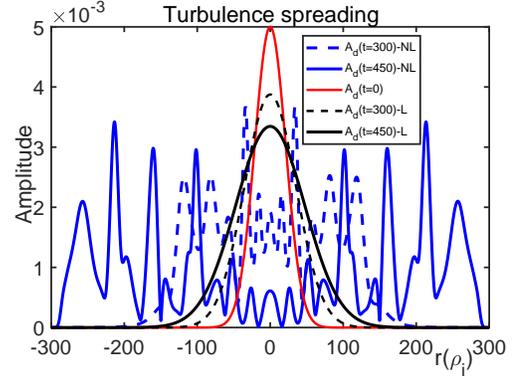}
\caption{The DW envelope at $t=0,300,450\omega_{*}^{-1}$ for both nonlinear
and linear cases. Here, the red solid curve is the initial DW structure, blue solid/dashed curves are nonlinear DW structures at $t=450\omega_*^{-1}$ and $t=300\omega_*^{-1}$, while black curves are their correspondence in the linear case.\label{fig:10}}
\end{figure}

To further verify whether  the split  structures are solitons, two initially
Gaussian DW envelopes are loaded, which then evolve and collide at about $t=250\omega^{-1}_*$, as shown in Fig. \ref{fig:9}.
However, their amplitude and shape do not change after the collision,
which is a significant feature of solitons.  The solitary structures are able to enhance turbulence spreading.
As shown in Fig. \ref{fig:10}, the nonlinearly generated solitons
propagate to a much broader radial extent than the linear dispersion,
thus enhancing turbulence spreading from linearly unstable to stable region. The radial propagation of  the coupled DW-GAM soliton exhibits convective nature, with $r\propto t$, which is much faster than any diffusive process, with $r\propto \sqrt{t}$. As a consequence, the nonlocal transport, a meso-scale phenomena,
may occur and influence the original local turbulence transport, which
may be the reason for the drastic transition of transport size scaling.

\section{Conclusion and discussion\label{sec:5}}

In this paper, the two-field model, governing nonlinear interaction
between DW and GAM, is derived and investigated in nonlinear gyrokinetic framework.
The numerical solution of the two-field model exhibits two phases with distinctive features, i.e., the linear growth stage and nonlinear saturated stage, with the former being extensively investigated by the well-studied PDI. This paper investigates both stages, while emphasizes the results in the latter. Furthermore, the influence of the DW-GAM solitons on the turbulence spreading from linearly unstable to stable region and its relation to the nonlocal transport and core-edge interaction are also discussed.

In the linear growth stage, the main results of the well-studied PDI, i.e., matching conditions for the resonant decay, the nonlinear growth rate of GAM and enhanced GAM radial propagation, are reproduced. Furthermore, in the nonlinear saturated stage, the initially spatially smooth DW envelope splits into multiple convectively propagating soliton structures, which are observed to spread to a much broader radial extent than the linear dispersion, exhibiting a convective propagation ($r\sim t$), instead of the typical dispersive behavior ($r\sim \sqrt{t}$) of linear propagation. The existence of soliton structures are further verified by the ``collision'' of the  structures formed by two initially loaded DW envelopes, which are observed to preserve their structures after the collision. The theoretical treatment yields a nonlinear schodinger equation describing the nonlinear evolution of DW due to GAM excitation, clearly indicating the competition between the quadratic DW kinetic dispersiveness and nonlinear trapping effect induced by GAM.

The theoretical analysis of the two-field model is not treated in a formal manner, which, however, provides a comprehensive
model to investigate fully nonlinear DW-GAM system, while though only the basic and preliminary aspect of the model  has been illuminated in the present work. Many important issues remain to be clarified. On the one hand, the DW and GAM should both be excited self-consistently from noise level under relevant profiles. On the other hand, the system nonuniformity, introduced by the nonuniformity of the diamagnetic drift frequency $\omega_*(r)$, may have even more significant effect on the nonlinear evolution than that in the PDI \cite{ZQiuPoP2014}, and is expected to affect the nonlocality of turbulence intensity. These aspects will be investigated in a future publication.

\section*{Acknowledgements}

This work is supported by   the National Key R\&D Program of China  under Grant No. 2017YFE0301900,
and the National Natural Science Foundation of China under grant No.  11875233. The authors, acknowledge Prof. Liu Chen (ZJU and UCI) and Dr. Fulvio Zonca (ENEA, Italy) for fruitful discussion, and Dr. Chen Zhao (ZJU and PPPL) for the help in producing Fig. 6.

\bibliography{reference}

\end{document}